\documentclass[aps,pra,showpacs,twocolumn]{revtex4}

\usepackage{graphicx}
\usepackage{dcolumn}
\usepackage{bm}

\usepackage{hyperref}
\hypersetup{colorlinks=true,linkcolor=red,citecolor = blue, urlcolor=black}

\usepackage[utf8]{inputenc}
\usepackage{amssymb}
\usepackage{amsthm}
\usepackage{mathrsfs}
\usepackage{dsfont}
\usepackage{multirow}
\usepackage{ upgreek }
\usepackage{enumitem}
\usepackage{ stmaryrd } 
\usepackage{float}
\usepackage{physics}
\usepackage{cancel}
\usepackage{subfig}

\newcommand{\ie}{\textit{i.e.}~}

\begin{document}
\title{Time-frequency metrology with two single-photon states: phase space picture and the Hong-Ou-Mandel interferometer}

\author{\'Eloi Descamps$^{1,2}$, Arne Keller$^{2,3}$, P\'erola Milman$^2$}
\affiliation{$^1$Département de Physique de l’\'Ecole Normale Supérieure - PSL, 45 rue d’Ulm, 75005 Paris, France}

\affiliation{$^2$Universitée Paris Cité, CNRS, Laboratoire Matériaux et Phénomènes Quantiques, 75013 Paris, France}

\affiliation{$^3$Département de Physique, Université Paris-Saclay, 91405 Orsay Cedex, France}

\date{\today}

\begin{abstract}
   We use time-frequency continuous variables as the standard framework to describe states of light in the subspace of individual photons occupying distinguishable auxiliary modes. We adapt to this setting the interplay between metrological properties and the phase space picture already extensively studied for quadrature variables. We also discuss in details the Hong-Ou-Mandel interferometer, which was previously shown to saturate precision limits, and provide a general formula for the coincidence probability of a generalized version of this experiment. From the obtained expression, we systematically analyze the optimality of this measurement setting for arbitrary unitary transformations applied to each one of the input photons. As concrete examples, we discuss transformations which can be represented as translations and rotations in time-frequency phase space for some specific states.
\end{abstract}

\maketitle

\section{Introduction}
Much has been discovered since the first proposals to use quantum systems in metrology.  From the role of entanglement \cite{RevModPhys.90.035006, RevModPhys.90.035005, Science, PhysRevA.95.032330} to the one of modes, for pure and noisy systems and measurements, several main results have been established, and the most important one is the fact that quantum mechanical protocols can provide a better scaling in precision with the number of probes than classical ones. Nevertheless, much still remains to be done, in particular concerning the application and the adaptation of such results to specific physical configurations. Of practical importance, for instance, is the issue of finding measurement strategies that lead to the optimal calculated limits, and this is far from being obvious for general states. Another relevant problem concerns adapting the general principles to physical constraints, as energy or temperature limits and thresholds \cite{safranek_optimal_2016,PhysRevA.99.043815}. Those are the main issues of this paper: in one hand, we deeply study the conditions for optimality of a specific measurement set-up and on the  other hand, we consider a specific physical system, consisting of individual photons, for measuring time and frequency related parameters.

In order to measure a given parameter $\kappa$ one performs an experiment producing different outcomes $x$ with associated probabilities $P_\kappa(x)$ and build an unbiased estimator $K$ such that  $\kappa=\expval{K}_\kappa$ is recovered. Here the index $\kappa$ means that we take the average for the probability distribution $P_\kappa$.  The Cramér-Rao bound (CRB) \cite{cramer_mathematical_2016} imposes a limit on the precision of parameter estimation:
\begin{equation}
    \delta\kappa\geq \frac{1}{\sqrt{N\mathcal{F}}},
\end{equation}
where, $\delta\kappa$ is the standard deviation in the estimation of $\kappa$: $\delta\kappa=\sqrt{\operatorname{Var}_\kappa(K)}$, $N$ is the number of independent measurements which were performed to estimate $\kappa$ and $\mathcal{F}$ is the quantity known as the Fisher information (FI), defined by : $\mathcal{F}=\int dx \frac{1}{P_{\kappa}(x)}\left(\frac{\partial P_{\kappa}(x)}{\partial\kappa}\right)^2 $.

In a quantum setting, one can use as a probe a quantum state $\ket{\psi}$ which can evolve under the action of an operator $\hat U(\kappa)=e^{-i\kappa\hat H}$ generated by an Hamiltonian $\hat H$. By optimizing the precision over all possible quantum measurements of a parameter $\kappa$, one obtains a bound, called the quantum Cramér-Rao bound (QCRB) \cite{braunstein_generalized_1996} which reads:
\begin{equation}
    \delta\kappa\geq \frac{1}{\sqrt{N\mathcal{Q}}},
\end{equation}
where $\mathcal{Q}$ is a quantity known as the quantum Fisher information (QFI) which for pure states and unitary evolutions (as the ones considered in the present paper), is equal to $\mathcal{Q}=4(\Delta\hat{H})^2$, with  $(\Delta\hat{H})^2=\bra{\psi(\kappa)}\hat{H}^2\ket{\psi(\kappa)}-\bra{\psi(\kappa)}\hat{H}\ket{\psi(\kappa)}^2$.\\

The FI indicates the precision of a given measurement, whereas the QFI is the maximum precision obtainable with any measurement. For a given setting, we can thus compute both quantities (FI and QFI) to have an idea if the measurement is optimal (QFI$=$FI) or not (QFI$>$FI).\\

Determining the QFI is a mathematical task much easier than finding a physical experimental set-up that reaches it. In quantum optical systems, several proposals and implementations exist where the QFI is indeed achieved \cite{PhysRevA.95.032330, PhysRevA.95.012104, SteveGear, Yu.18}, and one example where this is possible is the Hong-Ou-Mandel (HOM) experiment \cite{chen_hong-ou-mandel_2019, PhysRevA.106.063715, fabre_parameter_2021,lyons_attosecond-resolution_2018}. In this experiment, one focus on simple physical systems composed of two photons occupying distinguishable spatial modes with a given spectral distributions. This state is a particular example of a state defined in the single photon subspace (where each mode is populated by at most one photon), in which a general pure state that can be expanded as:
 \begin{equation}\label{state}
    \ket{\psi}=\int  d\omega_1\cdots  d\omega_n F(\omega_1,\cdots,\omega_n)\ket{\omega_1,\cdots,\omega_n}.
\end{equation}
In this formula, the indexes 1,2, ..$n$, label different auxiliary degrees of freedom (as for instance polarization or the propagation direction). The state $\ket{\omega_1,\cdots,\omega_n}$ is a pure state where each photon propagating in the mode $\alpha$ is exactly at the frequency $\omega_\alpha$. The spectral function $F$ also known as the joint spectral amplitude (JSA) is normalized to one: $\int \abs{F(\omega_1,...,\omega_n)}^2d\omega_1 ... d\omega_n=1$.

In this setting one can introduce time and frequency operators for each mode $\alpha$:  $\hat \omega_\alpha$ and $\hat t_\alpha$. They correspond respectively to the generators of time and frequency shifts of the photon in the mode labeled by $\alpha$.  An important property of these operators is that, in the considered single photon subspace  they satisfy the commutation relation $[\hat\omega_\alpha,\hat t_\beta]=i\delta_{\alpha,\beta}$ analogous to the one observed for the quadrature operators $\hat X_\alpha$ and $\hat P_\alpha$. Notice that we are using throughout this paper dimensionless operators, which are relative to particular time and frequency scales of the associated implementation. For a more complete description of the time frequency continuous variables one can refer to Appendix~\ref{annex: tf formalism} and to \cite{fabre_time_2022}.\\

Previous works on quantum metrology using the electromagnetic field quadratures or particles' position and momentum have shown how the phase space $(x_1,\cdots,x_n,p_1,\cdots,p_n)$ can provide not only insight but also an elegant geometrical picture of  the measurement precision \cite{fabre_time_2022,toscano_sub-planck_2006,zurek_sub-planck_2001}. Indeed the QFI can also be defined in terms of the Bures distance \cite{braunstein_statistical_1994} $s(\ket{\psi(\kappa)} , \ket{\psi(\kappa+d\kappa)})$: $\mathcal{Q}=4(\frac{s(\ket{\psi(\kappa)} , \ket{\psi(\kappa+d\kappa)})}{d\kappa})^2$. In the case of pure states, this distance is simply expressed in terms of the overlap $s(\ket{\psi} , \ket{\phi})= \sqrt{2(1-\abs{\braket{\phi}{\psi}})}$. Since the overlap of two states can be computed as the overlap of their respective Wigner function, one can interpret the QFI as a measure of how much the Wigner function must be shifted so as it becomes orthogonal to the initial one. A consequence of this is that the maximum precision of a measurement can be seen geometrically on the Wigner function, by looking at their typical size of variation in the direction of an evolution \cite{toscano_sub-planck_2006}. Since in the case of single photon states one can also define a time-frequency phase space associated to the variables $(\tau_1,\cdots,\tau_n,\varphi_1,\cdots,\varphi_n)$, it is natural to investigate wether the same type of interpretation makes sense in this context. 

The present paper purposes are thus twofold: in the first place, we provide general conditions for the HOM to saturate precision limits using time-frequency (TF) variables. For such, we consider arbitrary evolution operators acting on TF variables of single photons. In second place, we provide a phase-space picture and interpretation of the QFI for this type of system. Indeed, as shown in \cite{PhysRevA.105.052429}, there is an analogy between the quadrature phase space and the TF phase space from which metrological properties of time and frequency states can be inferred. Nevertheless, in the present case, photons have both spectral classical wave-like properties and quantum particle-like ones. Interpreting from a quantum perspective both the role of the spectral distribution and of collective quantum properties as entanglement in the single photon subspace has shown to demand taking a different perspective on the TF phase space \cite{descamps_quantum_2022}. Having this in mind, we investigate how relevant examples of  evolution operators, taken from the universal set of continuous variables quantum gates, can be implemented and represented in phase space, as well as the precision reached when one measures them using the HOM experiment. We'll concentrate on single-mode Gaussian operations, analogously to what was done in \cite{safranek_optimal_2016}, even though we provide a general formula for any transformation. 

This paper is organized as follows: In Section II we provide a description of the TF phase space and introduce the states we'll discuss in details as well as their representation.  In Section III we discuss the HOM experiment and the conditions for it to reach optimal precision limits. Finally, in Sections IV and V we discuss two different Gaussian operations in phase space as well as their implementation and the associated precision reached in the HOM experiment.  \\

\section{Time frequency phase space}\label{section: time freq phase space}

We consider pure two-photon states which can be written in the form: $\ket{\psi}=\int d\omega_1 d\omega_2 F(\omega_1,\omega_2)\ket{\omega_1,\omega_2}$. The Wigner function in variables $(\tau_1,\tau_2,\varphi_1,\varphi_2)$ of such states can be defined as
 \begin{widetext}
\begin{equation}\label{eq: Wigner function pure state}
    W_{\ket{\psi}}(\tau_1,\tau_2,\varphi_1,\varphi_2)=\int d\omega_1 d\omega_2 e^{2i(\omega_1\tau_1+\omega_2\tau_2)} F(\varphi_1+\omega_1,\varphi_2+\omega_2)F^\ast(\varphi_1-\omega_1, \varphi_2-\omega_2).
\end{equation}
\end{widetext}
Evolutions generated by $\hat\omega_\alpha$ and $\hat t_\alpha$ ($\alpha=1,2$) correspond to translations in phase space:
\begin{subequations}
\begin{align}
    W_{e^{-i\hat\omega_1\kappa}\ket{\psi}}(\tau_1,\tau_2,\varphi_1,\varphi_2)&=W_{\ket{\psi}}(\tau_1-\kappa,\tau_2,\varphi_1,\varphi_2),\\
    W_{e^{-i\hat t_1\kappa}\ket{\psi}}(\tau_1,\tau_2,\varphi_1,\varphi_2)&=W_{\ket{\psi}}(\tau_1,\tau_2,\varphi_1-\kappa,\varphi_2),
\end{align}
\end{subequations}
and analogously for $\hat\omega_2$ and $\hat t_2$.\\

Using the QFI formulation based on the Bures distance, we can safely state that the precision of a measurement device is related to its capability of distinguishing between an initial state $\ket{\psi(\kappa)}$ and a state $\ket{\psi(\kappa+d\kappa)}$ that has evolved according to a parameter $\kappa$. This precision is then directly related to how small the parameter $d\kappa$ should be such that these two states can be distinguished \ie the overlap $\abs{\bra{\psi(\kappa)}\ket{\psi(\kappa+d\kappa)}}$ gets close to zero. This can be also elegantly interpreted using the overlap of the two states's respective Wigner functions, that describe trajectories in the phase space that are governed by the interaction Hamiltonian and the parameter $d\kappa$.\\

To gain some familiarity with the studied problem we start with the case of a single-photon state $\ket{\psi}=\int d\omega S(\omega)\ket{\omega}$. Although using this type of state is not current in metrology, this simpler case can be seen as a building block and will help understanding the role of the spectrum in the present configuration.

For a single photon, the Wigner function is defined as: $W(\tau,\varphi)=\int d\omega e^{2i\omega\tau}S(\varphi+\omega)S^\ast(\varphi-\omega)$. In the case of a Gaussian state $\ket{\psi_G}$  with spectral wave function $S_G(\omega)=\frac{e^{-\frac{\omega^2}{4\sigma^2}}}{(2\pi\sigma^2)^{1/4}}$ its Wigner function is also Gaussian: $W_G(\tau,\varphi)= \exp(-2\sigma^2\tau^2-\frac{\varphi^2}{2\sigma^2})$. It is characterized by its width in the orthogonal directions $\tau$ and $\varphi$: $1/2\sigma$ and $\sigma$ respectively.\\

An evolution generated by $\hat\omega$ corresponds to a translation in the direction $\tau$ in phase space. The associated measurement precision is given by the smallest value of $d\kappa$ such that the initial Wigner function is almost orthogonal to the translated one in the corresponding direction. Since the width of the Wigner function in the direction of evolution is proportional to $1/\sigma$, we have $d\kappa\sim 1/\sigma$ leading to a QFI of the order of $\mathcal{Q}\sim \sigma^2$. Alternatively if one considers the generator $\hat t$, the associated width of the state will be $\sigma$ leading to a QFI of the order of $\mathcal{Q}\sim 1/\sigma^2$. We thus remark that the estimated QFI depends on the width of the state in phase space in the direction of evolution. We notice as well the similarities and differences with the quadrature phase space case: even though the relation between the phase space geometrical properties and metrological interest are common to both variables, in the case of quadrature they are related to some absolute quantum resource dependent quantity, the number of photons of the state. In the present case, the single photon spectrum is a classical resource and its width can only set a relative size scale in phase space. 

It is interesting to notice that this type of interpretation is also possible for classical fields, as studied in \cite{WalmsleySubPlanck,praxmeyer_time-frequency_2007,praxmeyer_direct_2016}. In this classical context, the electromagnetic field amplitude replaces the function $F$ and one can also relate spectral metrological properties to the phase space structures. Nevertheless, as discussed in \cite{descamps_quantum_2022}, this picture is merely associated to classical metrological properties of single mode fields (their spectrum) and no interesting scaling can be observed in this context. As a matter of fact, the classical single mode field and the single photon phase space can be mapped into one another. 

In the present paper, the multi-modal character of the quantum field is an essential ingredient for the discussion of the quantum metrological advantage, since it is a consequence of the multi-photon state. We will see in particular how these two features (spectral and particle-like) of the considered single photon subspace are combined in the QFI.

The situation is different and richer for bi-photon states, since the phase space is of dimension $4$. One can thus imagine different directions of translation  as for instance the ones generated by operators $\hat \omega_1$, $\hat \omega_2$, $\hat \omega_1-\hat \omega_2,\dots$ Then, optimizing the measurement precision involves, for a given spectral distribution, choosing a direction of evolution for which the Wigner function of the state has the smallest scale structures. This direction, as we'll see, will depend on the number of photons, and can display a non-classical scaling.

\section{The HOM as a measurement device}
\subsection{The setup}
\begin{figure}[h]
    \centering
    \includegraphics[width=0.8\linewidth]{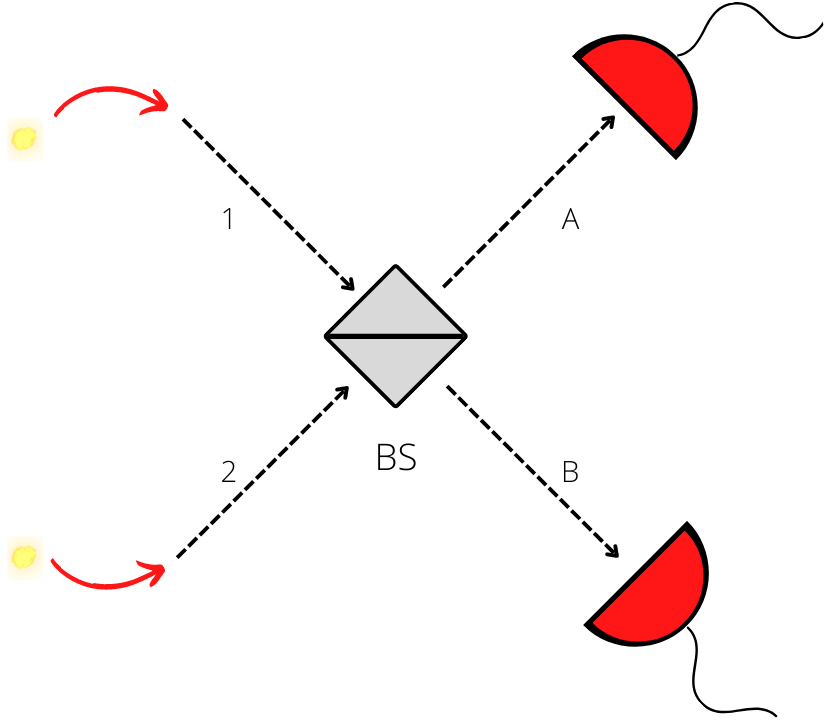}
    \caption{Schematic representation of HOM experiment.}
    \label{fig: HOM}
\end{figure}

In the setup proposed by Hong, Ou and Mandel \cite{hong_measurement_1987} two photons impinge into a balanced beam splitter (BS), each one of them from a different port, as represented on figure~\ref{fig: HOM}. By measuring the output of the beam-splitter using single-photon detectors we can compute the probability of obtaining coincidences (when the two photons exit the BS by different paths) or anti-coincidences (when they bunch and exit the BS at the same path). 

Since its original proposal and implementation, many modifications and adaptations were made to the HOM set-up, which was  shown to be very versatile to reveal different aspects of quantum optics using two-photon interference \cite{HOMEPJD}: it can be used to witness particle \cite{PhysRevA.83.042318} and spectral \cite{eckstein_broadband_2008} entanglement, to saturate precision bounds on time delay measurements\cite{chen_hong-ou-mandel_2019,PhysRevA.106.063715} or to directly measure the Wigner function of the incoming state \cite{douce_direct_2013,fabre_hongoumandel_2022}. 

We're interested in quantum metrological tasks, so we'll start by discussing the results obtained in \cite{chen_hong-ou-mandel_2019}, where the authors provided experimental evidence that the HOM device can saturate precision limits on time measurements. To achieve this result, the authors considered the initial state:
\begin{align}\label{eq: state ursin}
    \ket{\psi_U}&=\frac{1}{\sqrt{2}}\int d\Omega f(\Omega)\Big[\ket{\omega_1^0+\Omega,\omega_2^0-\Omega}-\notag\\
    &~~~~~~~~~~\ket{\omega_2^0+\Omega,\omega_1^0-\Omega}\Big],
\end{align}
where $\omega_1^0$ and $\omega_2^0$ are the central frequencies of the photons. Due to the energy conservation and to the phase-matching conditions, the support of the JSA associated to \eqref{eq: state ursin} is the line $\omega_1+\omega_2 =0$ in the plane $(\omega_1,\omega_2)$. It is anti-diagonal in the plane $(\omega_1,\omega_2)$ and infinitely thin along the diagonal direction $\omega_-=\omega_1-\omega_2$. Adding a delay in the arm $1$ of the HOM interferometer corresponds to an evolution generated by the operator $\hat\omega_1$, corresponding to a translation $\kappa$ in the $\tau_1$ direction. The QFI is simply calculated as: $\mathcal{Q}=4\Delta(\hat\omega_1)^2$. After the beam-splitter, the measurement can lead to two outcomes: coincidence or anti-coincidence, with probability $P_c$ and $P_a$, respectively. The FI is thus expressed as: $\mathcal{F}=\frac{1}{P_c}\left(\frac{\partial P_c}{\partial \kappa}\right)^2+\frac{1}{P_a}\left(\frac{\partial P_a}{\partial \kappa}\right)^2$. The authors of \cite{chen_hong-ou-mandel_2019} thus showed that using the input state \eqref{eq: state ursin} in the HOM interferometer, the two quantities $\mathcal{F}$ and $\mathcal{Q}$ are the same. 

In \cite{PhysRevA.106.063715} the HOM interferometer was also used and shown to lead to the QFI in a two-parameter estimation experiment. Finally, in \cite{fabre_parameter_2021} biphoton states were classified as metrological resources according to their spectral width, still in the situation where the HOM experiment is used as a measurement apparatus.

\subsection{Generalization: the HOM as an optimal measurement device for quantum metrology with biphotons}

We now make a general description of the HOM experiment as a parameter estimation device and try to understand and determine when it corresponds to an optimal measurement strategy. In \cite{PhysRevA.106.063715}, the authors tackle a part of this problem by studying the HOM as a measurement apparatus for two parameter estimation by establishing conditions on frequency correlation states. In this reference, the authors restrict themselves to time delay evolutions. 

\begin{figure}[h]
    \centering
    \includegraphics[width=0.8\linewidth]{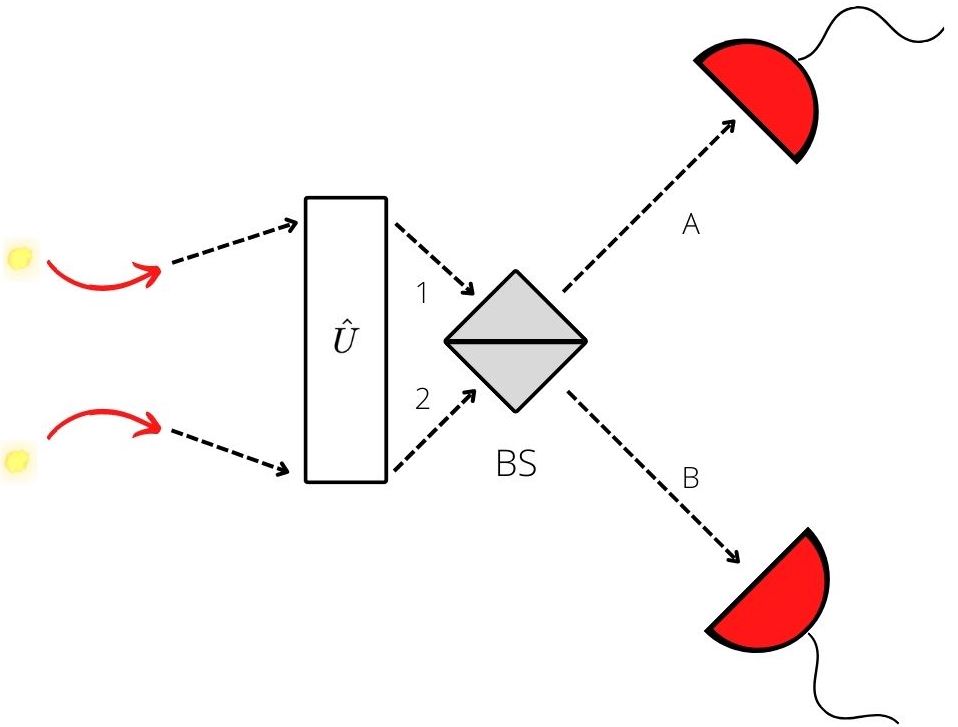}
    \caption{HOM setup where we apply a general gate $\hat{U}$ before the BS.}
    \label{fig: HOM with operator}
\end{figure}

In the present paper, we are interested in studying any evolution that can be described by a two photon unitary $\ket{\psi(\kappa)}=\hat U(\kappa)\ket{\psi}=e^{-i\hat H\kappa}\ket{\psi}$ (see figure~\ref{fig: HOM with operator}). We will see that under a symmetry assumption on the JSA of the state, it is possible to obtain an explicit formula for the FI, and this formula can be used to compute at a glance if the measurement setup considered is optimal or not.

For any input state $\ket{\psi}$, the QFI will then be expressed as:
\begin{equation}\label{eq: qfi}
    \mathcal{Q}=4\Delta(\hat H)^2.
\end{equation}
On the other hand, one can show that the coincidence probability is:
\begin{equation}\label{eq: pc}
    P_c=\frac{1}{2}(1-\bra{\psi}\hat U^\dagger \hat S \hat U \ket{\psi}).
\end{equation}
(see Appendix~\ref{annex: proofs}) where we introduced the hermitian swap operators $\hat S$ whose action on the states is given by $\hat S\ket{\omega_1,\omega_2}=\ket{\omega_2,\omega_1}$. Furthermore we can compute the associated FI. If the state $\ket{\psi}$ is symmetric or anti-symmetric (\textit{i.e.} $\hat S\ket{\psi}=\pm\ket{\psi}$) the FI at $\kappa=0$ it is given by:
\begin{equation}\label{eq: fi}
    \mathcal{F}=\Delta(\hat H-\hat S\hat H\hat S)^2.
\end{equation}
(see Appendix~\ref{annex: proofs}). This means that under the symmetry assumption on the JSA, comparing the QFI and the FI is done simply by comparing the variance of two different operators, mainly: $2\hat H$ and $\hat H-\hat S\hat H\hat S$. Equation~(\ref{eq: fi}) implies that if $[\hat H, \hat S]=0$, then $\mathcal{F}=0$ and no information can be obtained about $\kappa$ from the measurements. However, if $\{\hat H,\hat S\}=0$ then $\mathcal{F}=\mathcal{Q}$ since $\hat S\hat H\hat S=-\hat S^2\hat H=-\hat H$. In this last case, the measurement strategy is optimal. In \cite{PhysRevA.79.033822}, general conditions for reaching the QFI were also obtained in the context of  amplitude correlation measurements. These conditions are based on a quantum state's symmetry under (unphysical) path exchange.  \\

The previous calculations form a simple tool that can be applied to different evolution Hamiltonians $\hat H$. We'll now discuss examples taken from the universal set of quantum gates in continuous variables: translations (generated by operator  $\hat\omega_\alpha$'s) and rotations (generated by $\hat H=(\hat \omega^2+\hat t^2)/2$). These gates have already been studied in \cite{safranek_optimal_2016} in the case of quadrature or position and momentum. In the present physical configuration, they correspond to the free evolution of single photons in free space (translations) or in a dispersive medium, as for instance an optical fiber combined to time lenses (rotation).

\section{Time-frequency phase-space translations}
\subsection{Different types of translations}

Since we're considering two-photon states,  translations can be represented by any linear combination of the corresponding operators, that is : $\hat H=\alpha\hat \omega_1+\beta\hat \omega_2+\gamma\hat t_1+\delta\hat t_2$. To illustrate our results we choose to focus on the four operators $\hat \omega_1$, $\hat\omega_2$ and $\hat\omega_{\pm}= \hat \omega_1 \pm \hat \omega_2$, since they are the most easily implemented in HOM experiment. Notice that $\hat \omega_{\pm}$ are collective operators acting in both input photons while $\hat \omega_{1,2}$ act in a single photon only.\\

If we consider a state which is (anti-)symmetric and separable in the variables $\omega_\pm=\omega_1\pm\omega_2$, we can write:
\begin{equation}\label{eq: etat general}
    \ket{\psi}=\frac{1}{\sqrt{2}}\int d\omega_+d\omega_-f(\omega_+)g(\omega_-)\ket{\frac{\omega_++\omega_-}{2},\frac{\omega_+-\omega_-}{2}},
\end{equation}
with $g$ satisfying $g(-\omega)=\pm g(\omega)$ and the functions $g$ and $f$ being normalized to one. The specific form of each function is related to the phase-matching conditions and the energy conservation of the two-photon generation process and this type of state can be experimentally produced in many set-ups \cite{boucher_toolbox_2015, ramelow_discrete_2009}. Using equations (\ref{eq: qfi}) and (\ref{eq: fi}) we can compute the QFI and FI associated to each type of evolution:
\begin{itemize}
    \item For $\hat H=\hat \omega_1$, we get $\mathcal{Q}=\Delta(2\hat \omega_1)^2=\Delta(\hat \omega_++\hat \omega_-)^2=\Delta(\hat \omega_-)^2+\Delta(\hat\omega_+)^2$, while $\mathcal{F}=\Delta(\hat \omega_-)^2$. Thus this situation is optimal only if $\Delta(\hat \omega_+)^2=0$, which was the case for the state $\ket{\psi_U}$ of Eq. \eqref{eq: state ursin} used in \cite{chen_hong-ou-mandel_2019}). We obtain the same type of result for $\hat\omega_2$.
    \item For $\hat H=\hat \omega_+$, $\mathcal{Q}=4\Delta(\hat \omega_+)^2$, while $\mathcal{F}=\Delta(\hat \omega_+-\hat\omega_+)^2=0$. In this situation the precision of the measurement is zero, and the reason for that is that variables $\omega_+$ cannot be measured using the HOM experiment (we notice that $[\hat \omega_+,\hat S]=0$).
    \item For $\hat H=\hat \omega_-$, we get $\mathcal{Q}=4\Delta(\hat \omega_-)^2$, while $\mathcal{F}=\Delta(\hat \omega_-+\hat\omega_-)^2=4\Delta(\hat\omega_-)^2$. This time we have $\mathcal{F}=\mathcal{Q}$, which means that the measurement is optimal. In this case, we have that $\{\hat \omega_-,\hat S\}=0$.
\end{itemize}

We now illustrate these general expressions and interpret them using different quantum states and their phase space representations.

\subsection{Example: Gaussian and Schrödinger cat-like state}

To illustrate our point we discuss as an example two states $\ket{\psi_G}$ and $\ket{\psi_C}$ that can be expressed in the form of equation~(\ref{eq: etat general}). For $\ket{\psi_G}$, $f$ and $g$ are Gaussians:
\begin{align}
    f_G(\omega_+)=\frac{e^{-\frac{(\omega_+-\omega_p)^2}{4\sigma_+^2}}}{(2\pi\sigma_+^2)^{1/4}} && g_G(\omega_-)=\frac{e^{-\frac{\omega_-^2}{4\sigma_-^2}}}{(2\pi\sigma_-^2)^{1/4}},
\end{align}
where $\sigma_\pm$ is the width of the corresponding function and $\omega_p$ is a constant, which is also the photon's central frequency. As for state $\ket{\psi_C}$, it can be seen as the generalization of (\ref{eq: state ursin}). We consider $f$ to be Gaussian and $g$ to be the sum of two Gaussians:
\begin{align}
    &f_C(\omega_+)=f_G(\omega_+) \notag\\ &g_C(\omega_-)=\frac{1}{\sqrt{2}}\Big[g_G(\omega_-+\Delta/2)-g_G(\omega_--\Delta/2)\Big],
\end{align}
where $\Delta$ is the distance between the two Gaussian peaks of $g_C$. We assume that the two peaks are well separated: $\Delta\gg\sigma_-$. Consequently, $g_C$ is approximately normalized to one. We can verify that with these definitions the function $g_G$ is even while $g_C$ is odd by exchange of variables $\omega_1$ and $\omega_2$.
We first compute the variances for both states (table~\ref{tab: variance translation}) and then apply the formula (\ref{eq: qfi}) and (\ref{eq: fi}).

\begingroup
\setlength{\tabcolsep}{10pt} 
\renewcommand{\arraystretch}{1.5} 
\begin{table}[h]
    \centering
    \begin{tabular}{|c|c|c|}
        \hline
        State & $\ket{\psi_G}$ & $\ket{\psi_C}$ \\
        \hline
        \hline
        $\Delta(\hat\omega_1)^2$ or $\Delta(\hat\omega_2)^2$ &  $\frac{1}{4}\sigma_+^2+\frac{1}{4}\sigma_-^2$ & $\frac{1}{16}\Delta^2+\frac{1}{4}\sigma_+^2+\frac{1}{4}\sigma_-^2$\\
        \hline
        $(\Delta\hat\omega_+)^2$ & $\sigma_+^2$ & $\sigma_+^2$\\
        \hline
        $(\Delta\hat\omega_-)^2$ & $\sigma_-^2$ & $\frac{1}{4}\Delta^2+\sigma_-^2$\\
        \hline
\end{tabular}
    \caption{Variance of various time translation operators for states $\ket{\psi_G}$ and $\ket{\psi_C}$. See Appendix C for details.}
    \label{tab: variance translation}
\end{table}
\endgroup

So for the case of an evolution generated by $\hat \omega_1$, for $\ket{\psi_G}$ we obtain:
\begin{align}
    \mathcal{Q}=\sigma_+^2+\sigma_-^2 && \mathcal{F}=\sigma_-^2,
\end{align}
while for $\ket{\psi_C}$ we have:
\begin{align}
    \mathcal{Q}=\frac{1}{4}\Delta^2+\sigma_+^2+\sigma_-^2 && \mathcal{F}=\frac{1}{4}\Delta^2+\sigma_-^2.
\end{align}
We thus see that time precision using the HOM measurement and the quantum state evolution generated by $\hat\omega_1$ is optimal only if the parameter $\sigma_+$ is negligible compared to $\Delta$ or $\sigma_-$. This is exactly the case for the state~(\ref{eq: state ursin}) where $\sigma_+=0$. 

In addition, we see that there is a difference between the QFI associated to $\ket{\psi_C}$ and $\ket{\psi_G}$ involving the parameter $\Delta$. This difference can be interpreted, as discussed in  \cite{fabre_parameter_2021}, as a spectral effect. In this reference, the spectral width is considered as a resource, and for a same spectral width  state $\ket{\psi_C}$ has a larger variance than state $\ket{\psi_G}$. Nevertheless, as discussed in \cite{descamps_quantum_2022}, this effect has a classical spectral engineering origin and choosing to use one rather than the other depends on the experimentalists constraints.

\subsection{Interpretation of translations in the time-frequency phase space}

\begin{figure}
    \begin{tabular}{cc}
        \subfloat[Projection on the plane $\tau_-$, $\omega_-$] {\includegraphics[width=0.5\linewidth]{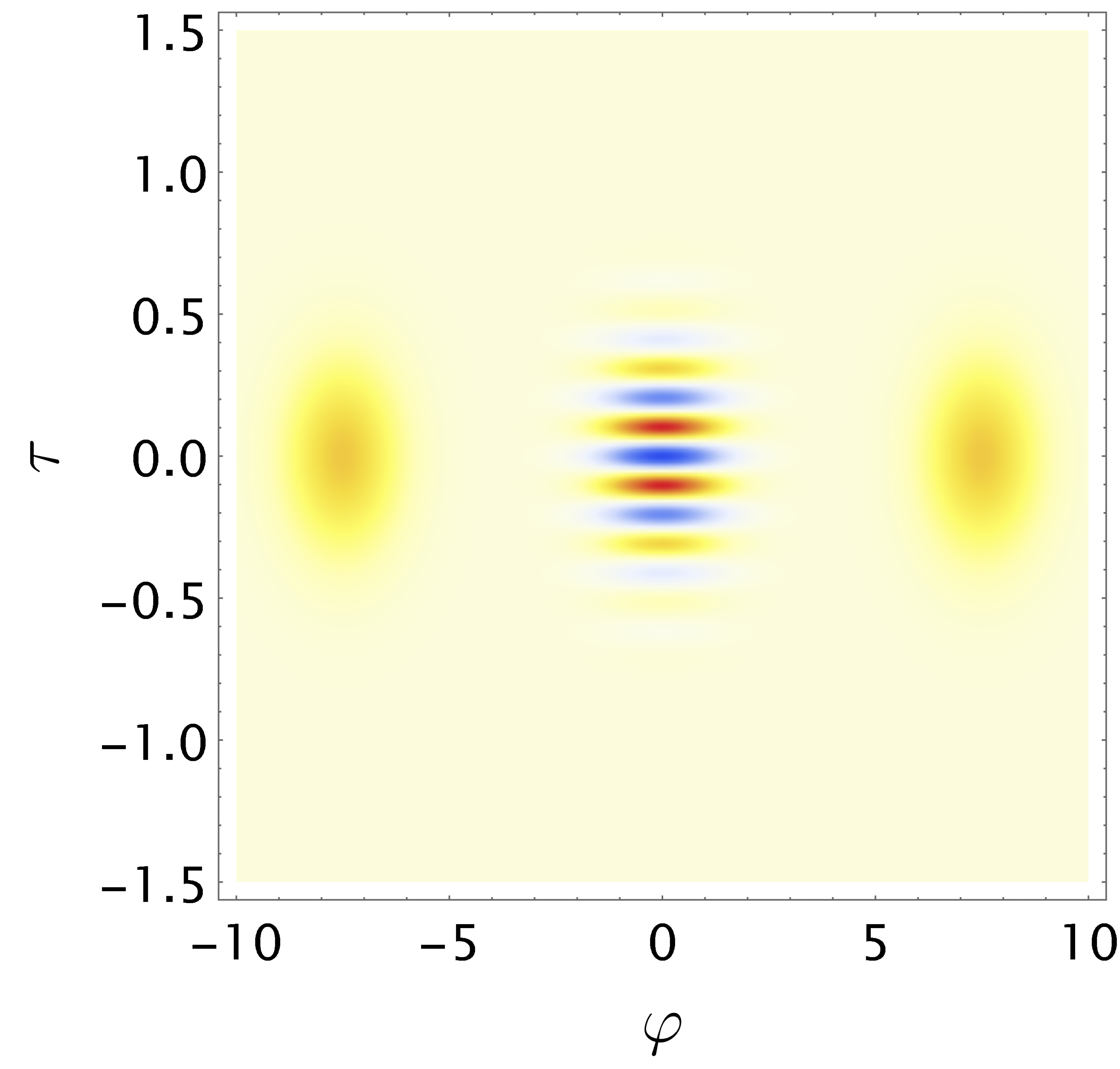}} 
        & \subfloat[Projection on the plane $\tau_1$, $\omega_1$] {\includegraphics[width=0.5\linewidth]{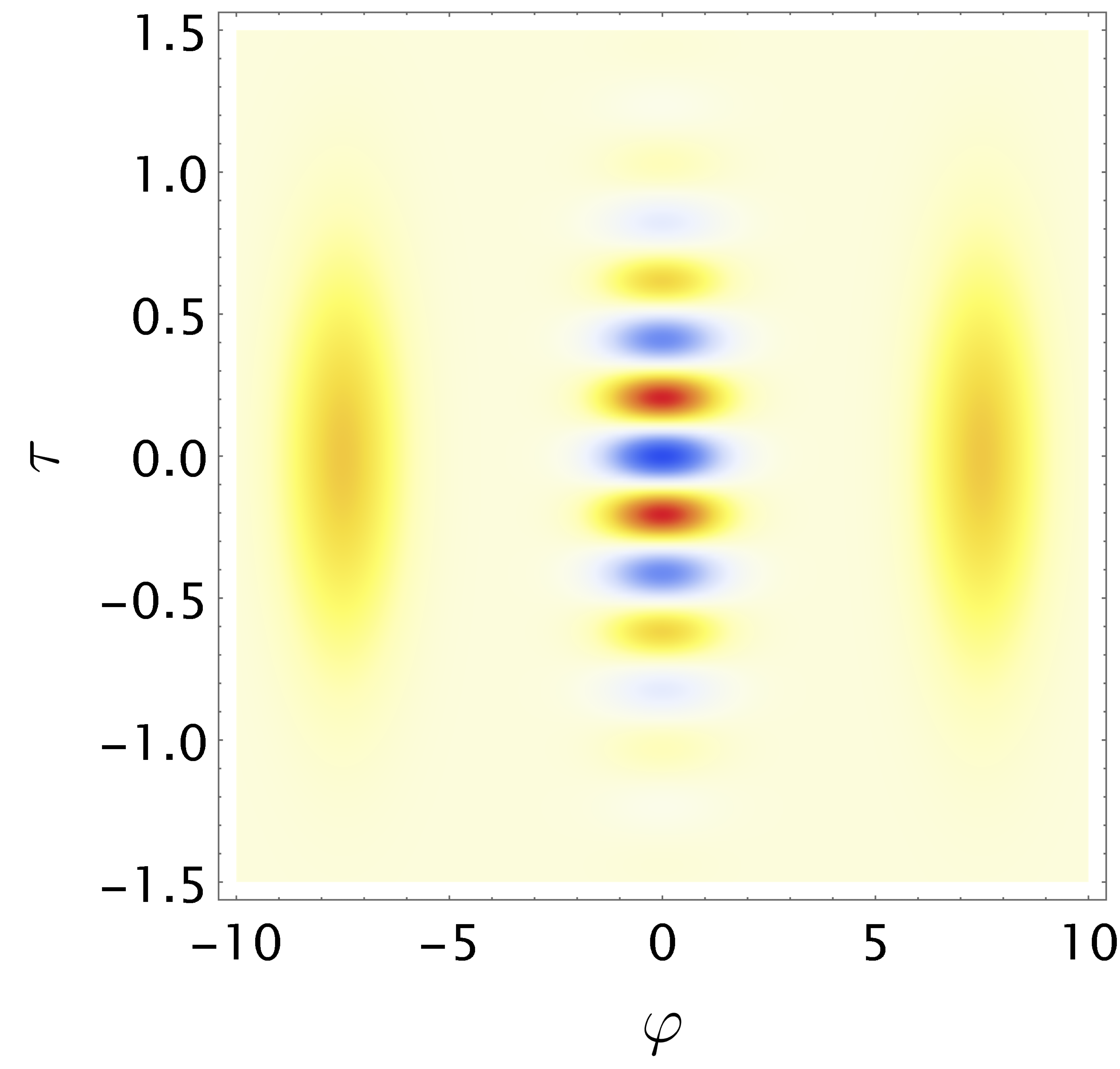}}
    \end{tabular}
    \caption{Wigner function of the cat-like state $\ket{\psi_C}$ projected in different variables.}\label{fig: wigner}
\end{figure}

We now discuss the dependency of precision on the direction of translation. For such, we can consider the Wigner function associated to a JSA which is separable in the $\omega_\pm$ variables. Its Wigner function will also be separable on these variables:
\begin{equation}\label{eq: factorization wigner function}
    W(\tau_1,\tau_2,\varphi_1,\varphi_2)=W_+(\tau_+,\varphi_+)W_-(\tau_-,\varphi_-),
\end{equation}
where the phase space variables $\tau_\pm$ and $\varphi_\pm$ are defined as: $\varphi_\pm=\frac{\varphi_1\pm\varphi_2}{2}$ and $\tau_\pm=\tau_1\pm\tau_2$. Even though the Wigner function $W_+$ (resp. $W_-$) can be associated to the one of a single variable ($\omega_+$ ($\omega_-$)) and spectral wave function $f$ (resp. $g$), it
displays some differences with the single photon one. This fact is well illustrated in Fig. ~\ref{fig: wigner}.

For state $\ket{\psi_C}$, according to \eqref{eq: factorization wigner function} the projection of the Wigner function $W_-$ in the plane $\tau_-,\phi_-$ of the phase space can be represented as show in Figure~\ref{fig: wigner} (a). We see that it is composed of two basic shapes: two Gaussian peaks and an oscillation pattern in between. Figure \ref{fig: wigner} (b) represents another way to project this very same Wigner function onto the plane $\tau_1, \phi_1$ of the phase space. One can observe that in this case the distance between the peaks is larger than in the previous representation by a factor of $2$. As precision is directly related to the size of the Wigner function structures in phase space, we observe that the interference fringes are closer apart in the phase space associated to the minus variable than in the one associated to mode 1. Thus, the precision in parameter estimation will be better using  $\hat \omega_-$ as the generator of the evolution than when using $\hat \omega_1$. This phase space based observations explain well the result of the computation of the QFI:
\begin{align}\label{eq: difference variance direction}
    4\Delta(\hat{\omega}_1)^2=\Delta(\hat{\omega}_-)^2.
\end{align}
with the assumption that $\sigma_+\ll\Delta,\sigma_-$.

The reason for the appearance of a factor $2$ difference in fringe spacing for the Wigner function associated to variable $\omega_-$ is the fact that it is a collective variable, and translations in the phase space associated to these variables are associated to collective operators, acting on both input photons (instead of a single one, as is the case of translations generated by operator $\hat \omega_1$, for instance). Thus, one can observe, depending on the biphoton quantum state ({\it i.e.}, for some types of frequency entangled states), a scaling depending on the number of particles (in this case, two). As analyzed in \cite{descamps_quantum_2022} for general single photon states composed of $n$ individual photons, we have for frequency separable states a scaling corresponding to the shot-noise one ({\it i.e.}, proportional to $\sqrt{n}$). A Heisenberg-like scaling (proportional to $n$) can be achieved for non-physical maximally frequency correlated states, and considering a physical non-singular spectrum leads to a non-classical scaling in between the shot-noise and the Heisenberg limit.  

Experimentally, such collective translation can be implemented by adding a delay of $\tau$ in arm 1 and of $-\tau$  in arm 2. Notice that this situation is different from creating a delay of $2 \tau$ in only one arm, even though both situations lead to the same experimental results in the particular context of the HOM experiment.

\section{Time-frequency phase space rotations}

We now move to the discussion of the phase space rotations. For this, we'll start by providing some intuition by discussing in first place the single photon (or single mode) situation. In this case, time-frequency phase space rotations are generated by the operators $\hat R=\frac{1}{2}(\hat\omega^2+\hat t^2)$. As previously mentioned, we consider here dimensionless observables. Physically, time-frequency phase space rotations correspond to performing a fractional Fourier transform of the JSA. While for transverse variable of single photons the free propagation or a combination of lenses can be used for implementing this type of operation \cite{tasca_continuous_2011, Pellat-Finet:94}, in the case of time and frequency this transformation corresponds to the free propagation in a dispersive medium \cite{doi:10.1063/5.0009527, TanzNPJ, PhysRevA.80.043821, TempFT, Lu:18} combined to temporal lenses \cite{Mazelanik:20, fabre_quantum_nodate, PhysRevLett.94.073601}.

\subsection{Single mode rotations}

In this Section, we compute the QFI associated to a rotation $\hat{R}$ for a single photon, single mode state using the variance of this operator for different states $\ket{\psi}=\int d\omega S(\omega)\ket{\omega}$. As for the translation, this simpler configuration is used as a tool to better understand the two photon case.\\

\subsubsection{Gaussian state:}

We start by discussing a single-photon Gaussian state at central frequency $\omega_0$ and spectral width $\sigma$:
\begin{equation}
    \ket{\psi_G(\omega_0)}=\frac{1}{(2\pi\sigma^2)^{1/4}}\int d\omega e^{-\frac{(\omega-\omega_0)^2}{4\sigma^2}}\ket{\omega}.
\end{equation}
For this state, we have that:
\begin{equation}\label{eq: Variance rotation Gaussian state}
    \Delta(\hat{R})^2=\sigma^2\omega_0^2+\frac{1}{8}\left[\frac{1}{4\sigma^4}+4\sigma^4-2\right].
\end{equation}
Eq. \eqref{eq: Variance rotation Gaussian state} has two types of contributions that we can interpret:
\begin{itemize}
    \item The first term $\sigma^2\omega_0^2$ corresponds to the distance in phase space ($\omega_0$) of the center of the distribution, to the origin of the phase space ($\omega=0, \tau=0$), times the width of the state $\sigma$ in the direction of rotation  (see Figure~\ref{fig: rotation states}~(a)).  This term is quite intuitive. The Wigner function of a state which is rotated by an angle $\theta=1/2\sigma\omega_0$ has an  overlap with the Wigner function of the initial one which is close to zero. 
    \item The term $\frac{1}{4\sigma^4}+4\sigma^4-2$ reaches $0$ as a minimum when $\sigma=\frac{1}{\sqrt{2}}$. For this value the Wigner function is perfectly rotationally symmetric. Its meaning can be intuitively understood if we consider that   $\omega_0=0$, so that this term becomes the only contribution to the variance(see Figure~\ref{fig: rotation states}~(b)). In this case, we are implementing a rotation around the center of the state. If the state is fully symmetric then this rotation has no effect, and the  variance is $0$. Only in the case where the distributions rotational symmetry is broken we obtain a non zero contribution.
\end{itemize}

\subsubsection{Schr\"odinger cat-like state centered at the origin ($\omega=0$):}

We now consider the superposition of two Gaussian states:
\begin{equation}
    \ket{\psi_C^0}=\frac{1}{\sqrt{2}}(\ket{\psi_G(\Delta/2)}-\ket{\psi_G(-\Delta/2)}).
\end{equation}
This state is of course non physical as a single-photon state, since it contains negative frequencies. However, since it can be be well defined using collective variables (as for instance $\omega_-$) for a two or more photons state, we still discuss it.
Assuming that the two peaks are well separated ($\Delta\gg \sigma$), we can ignore the terms proportional to $e^{-\frac{\Delta^2}{8\sigma^2}}$, and this leads to:
\begin{equation}\label{Variance rotation cat state center at 0}
    \Delta(\hat{R})^2=\frac{1}{8}\left[\frac{1}{4\sigma^4}+4\sigma^4-2\right]+\frac{1}{4}\Delta^2\sigma^2.
\end{equation}
We see that there is no clear metrological advantage when using this state compared to the Gaussian state: the quantity $\Delta/2$ plays the same role as $\omega_0$. This can be understood geometrically once again, with the help of the Wigner function. We see in Figure~\ref{fig: rotation states}~(c) how the considered state evolves under a rotation. In this situation the interference fringes are rotated around their center so even though they display a small scale structure, they are moved only by a small amount, resulting in a non significant precision improvement. \\

\subsubsection{Schr\"odinger cat-like state centered at any frequency:}

\begin{figure}
\begin{tabular}{cc}
\subfloat[Gaussian state centered at $\omega_0$. For $\theta\omega_0\sim 1/2\sigma$ the initial state and the rotate one are distinguishable.]{\includegraphics[width=0.5\linewidth]{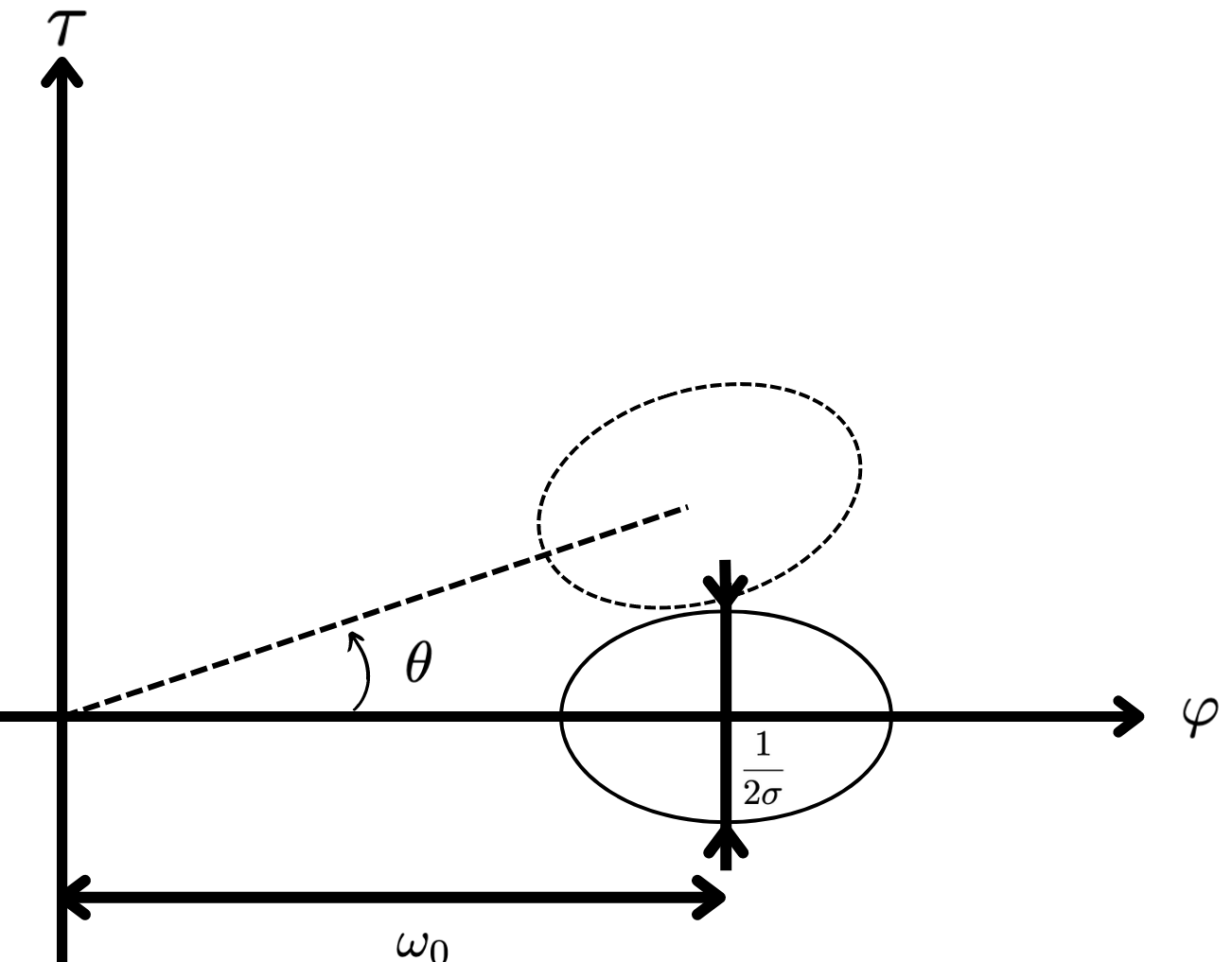}} 
   & \subfloat[Gaussian state centered at the origin. The rotated state will be distinguishable from the initial one only in the absence of rotational symmetry.]{\includegraphics[width=0.5\linewidth]{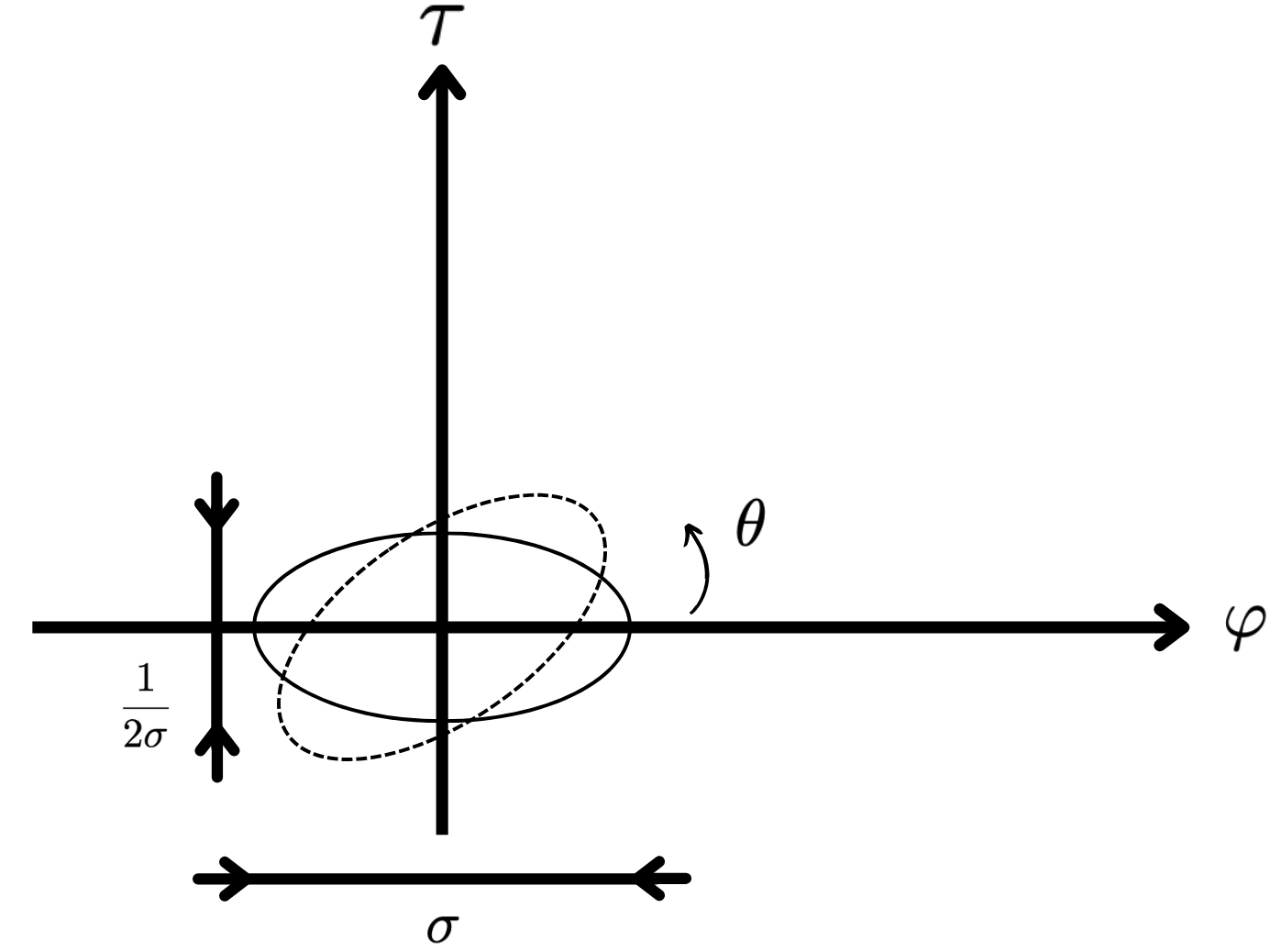}}\\
\subfloat[Superposition of two Gaussian states (cat-like state) centered the origin. The small structures of the fringes do not play a relevant role since they are only moved by a small distance under rotation.]{ \includegraphics[width=0.5\linewidth]{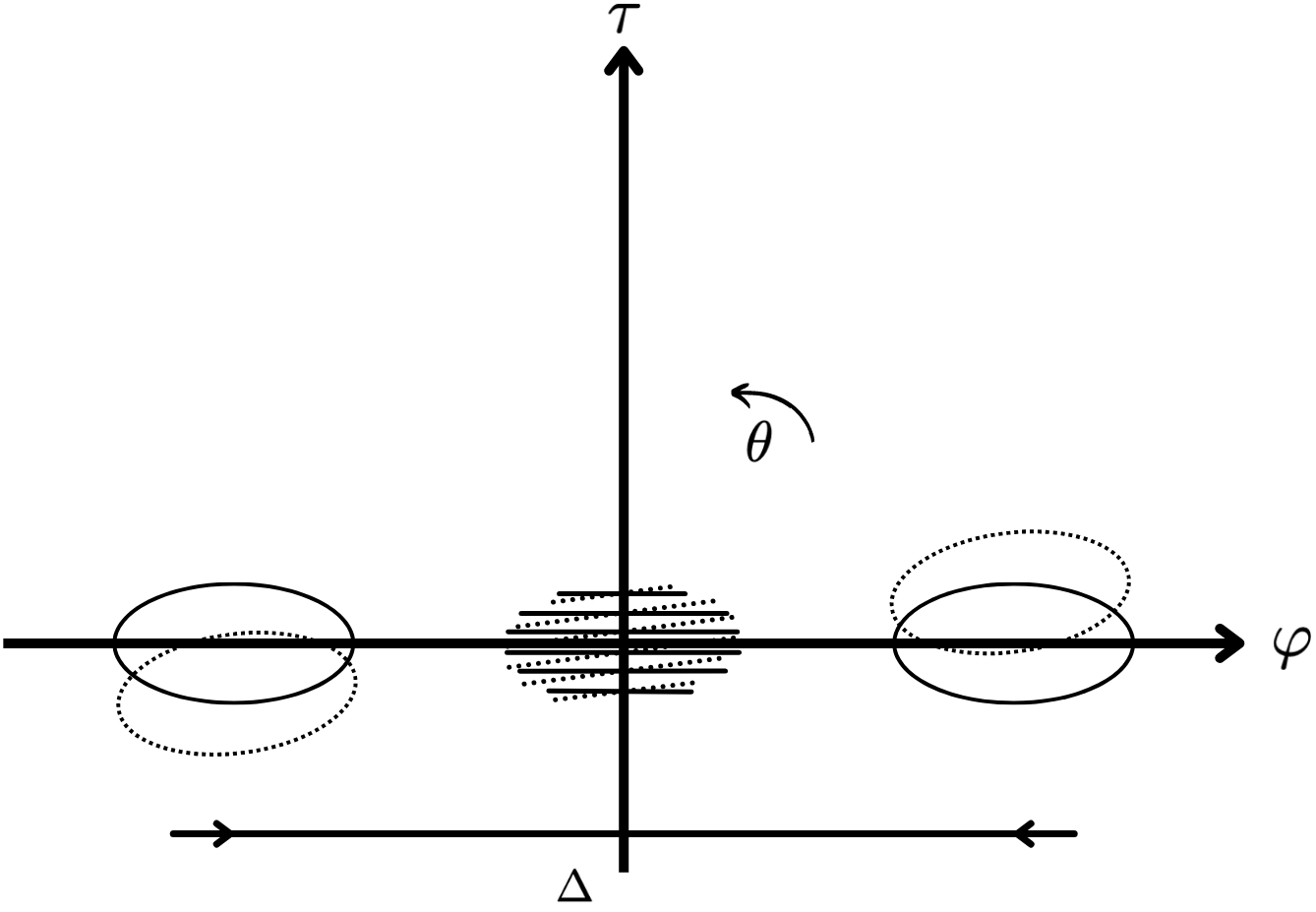}} 
   & \subfloat[Superposition of two Gaussian states (cat-like state) centered at $\omega_0$. The fringes play an important role, since with $\theta\omega_0\sim 1/\Delta$, the two states are nearly orthogonal.]{\includegraphics[width=0.5\linewidth]{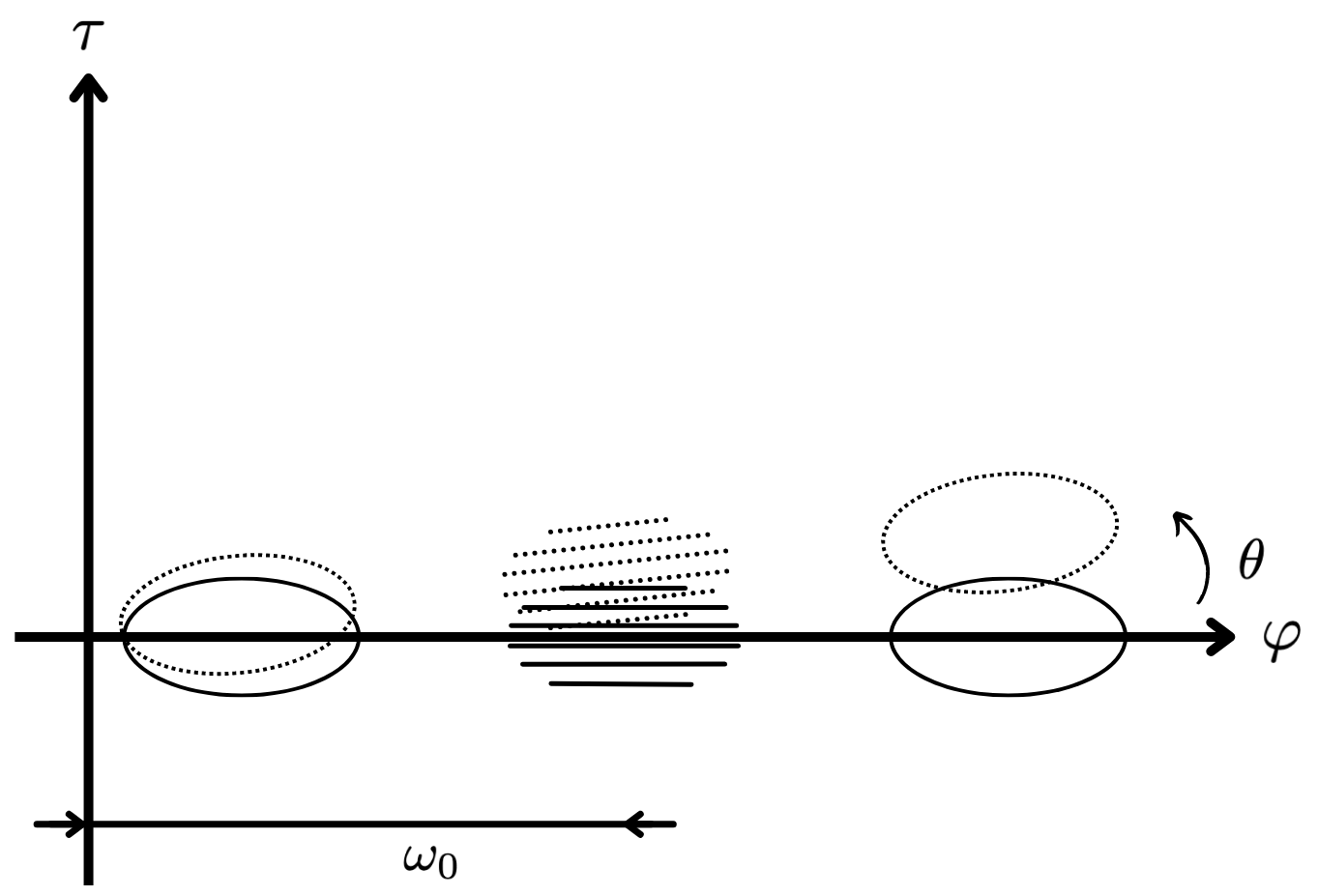}}\\

\end{tabular}
\caption{Schematic representation of the Wigner function of various states under rotation. The ellipses represent the typical width of Gaussians. The doted lines represent the rotated states.}\label{fig: rotation states}
\end{figure}

We can now discuss the state formed by the superposition of two Gaussian states whose peaks are at frequencies $\omega_0-\Delta/2$ and $\omega_0+\Delta/2$, and with the same spectral width $\sigma$ as previously considered:
\begin{equation}
    \ket{\psi_C}=\frac{1}{\sqrt{2}}\Big(\ket{\psi_G(\omega_0+\Delta/2)}-\ket{\psi_G(\omega_0-\Delta/2)}\Big).
\end{equation}
Still under the assumption of a large separation between the two central frequencies ($\Delta\gg\sigma$), we obtain:

\begin{equation}
    \Delta(\hat{R})^2=\frac{1}{8}\left[\frac{1}{4\sigma^4}+4\sigma^4-2\right]+\frac{1}{4}\Delta^2(\sigma^2+\omega_0^2)+\sigma^2\omega_0^2.
\end{equation}
We can notice that by setting $\omega_0=0$ we recover the variance corresponding to the same state rotated around its center. Nevertheless, in the present case $\omega_0 \neq 0$, and we have two additional terms: $\sigma^2\omega_0^2$ and $\Delta^2\omega_0^2/4$. Both terms can be interpreted as a product of the state's distance to the origin and its structure in phase space. However, while the first one is simply the one corresponding to the Gaussian state, the second one is a product of the states' distance to the origin and its small structures in phase space, created by the interference between the two Gaussian states (see Figure \ref{fig: rotation states} (d)). The interference pattern is thus rotated by an angle $\theta$ corresponding to an arc of length $\omega_0\theta$, and since the distance between the fringes is of order $\Delta$, if $\theta \sim 1/\omega_0\Delta$ (corresponding to the term $\Delta^2\omega_0^2/4$ in the expression of the variance) the rotated state is close to orthogonal to the initial one.

In all this section, we have considered rotations about the time and frequency origin of the phase space. Nevertheless, it is of course possible to displace this origin and consider instead rotations about different points of the TF phase space. In this case, for a rotation around an arbitrary point $\tau_0$ and $\varphi_0$, the generator would be given by $(\hat \omega-\varphi_0)^2/2 +(\hat t-\tau_0)^2/2$.

\subsection{Different types of rotations}

We now move to the case of two single photons (biphoton states). As for the case of translations, there are many possible variables and can consider rotations in different planes of the phase space: $\hat R_1$, $\hat R_2$, $\hat R_\pm$, $\hat R_1\pm \hat R_2 \dots$ where $\hat R_1=\frac{1}{2}(\hat\omega_1^2+\hat t_1^2)$ (and similarly for $\hat R_2$) and $\hat R_{\pm}=\frac{1}{4}(\hat\omega_\pm^2+\hat t_\pm^2)$ (recall that $\hat\omega_\pm=\hat\omega_1\pm\hat\omega_2$ and $\hat t_{\pm}=\hat t_1\pm \hat t_2$).
For all these operators we can as before apply the general formula for the QFI and of the FI to the corresponding HOM measurement. The results are displayed in table~\ref{tab: qfi fi rotation}.\\

\begingroup
\setlength{\tabcolsep}{10pt} 
\renewcommand{\arraystretch}{1.5} 
\begin{table}[H]
    \centering
    \begin{tabular}{|c|c|c|}
        \hline
        Operator & QFI & FI \\
        \hline
        \hline
        $\hat R_1$ & $4\Delta(\hat R_1)^2$ &  $\Delta(\hat R_1-\hat R_2)^2$ \\
        \hline
        $\hat R_\pm$ & $4\Delta(\hat R_\pm)^2$ & $0$\\
        \hline
        $\hat R_1+\hat R_2$ & $4\Delta(\hat R_1+\hat R_2)^2$ & $0$\\
        \hline
        $\hat R_1-\hat R_2$ & $4\Delta(\hat R_1-\hat R_2)^2$ & $4\Delta(\hat R_1-\hat R_2)^2$\\
        \hline
\end{tabular}
    \caption{QFI and FI of various rotation operators.}
    \label{tab: qfi fi rotation}
\end{table}
\endgroup

We see that the only two situations where the HOM can indeed be useful as a measurement device for metrological applications are $\hat R_1$ and $\hat R_1-\hat R_2$. The reason for that is the symmetry of $\hat R_\pm$ and $\hat R_1+\hat R_2$, which commute with the swap operator $\hat S$. As for $\hat R_1$, it corresponds to the rotation of only one of the photons and may not be the optimal strategy. Finally, $\hat R_1-\hat R_2$ corresponds to the simultaneous rotation in opposite directions of both photons sent into the two different input spatial modes. As $\hat R_1-\hat R_2$ anti-commutes with $\hat S$ then we can affirm that the HOM measurement is optimal for this type of evolution.

\subsection{QFI and FI computation with Gaussian and cat-like state}

We now compute the QFI and FI using the variance of $\hat R_1$ and $\hat R_1-\hat R_2$ calculated for states $\ket{\psi_G}$ and $\ket{\psi_C}$.\\

\underline{For $\ket{\psi_G}$:}\\
We have:
\begin{align}
    \Delta(\hat R_1)^2&=\frac{1}{32}\left[\left(\frac{1}{\sigma_+^2}+\frac{1}{\sigma_-^2}\right)^2+(\sigma_+^2+\sigma_-^2)^2-8\right]\notag\\
    &~~~~~~+\frac{1}{16}\omega_p^2(\sigma_+^2+\sigma_-^2)\notag\\
    \Delta(\hat R_1-\hat R_2)&=\frac{1}{4}\left[\frac{1}{\sigma_+^2\sigma_-^2}+\sigma_+^2\sigma_-^2-2\right]+\frac{1}{4}\sigma_-^2\omega_p^2.
\end{align}

\underline{For $\ket{\psi_C}$:}\\
We have:
\begin{align}
    \Delta(\hat R_1)^2&=\frac{1}{32}\left[\left(\frac{1}{\sigma_+^2}+\frac{1}{\sigma_-^2}\right)^2+(\sigma_+^2+\sigma_-^2)^2-8\right]\notag\\
    &~~~~~~+\frac{1}{64}(4\omega_p^2+\Delta^2)(\sigma_+^2+\sigma_-^2)\notag\\
    &~~~~~~+\frac{1}{64}\Delta^2\omega_p^2+\frac{\Delta^2}{128}\left(\frac{1}{\sigma_-^2}+\sigma_-^2\right)\notag\\
    \Delta(\hat R_1-\hat R_2)&=\frac{1}{4}\left[\frac{1}{\sigma_+^2\sigma_-^2}+\sigma_+^2\sigma_-^2-2\right]+\frac{1}{4}\sigma_-^2\omega_p^2.
\end{align}

We notice that for both states $4\Delta(\hat R_1)^2\geq \Delta(\hat R_1-\hat R_2)^2$, meaning that the measurement of a rotation implemented in only one mode using the HOM is not an optimal measurement.\\

Experimentally realizing an evolution generated by $\hat R_1$ is easier than implementing the one associated to $\hat R_1-\hat R_2$. Furthermore we see that for the Gaussian state $\ket{\psi_G}$ a dominant term is $\omega_p^2\sigma_-^2$ which appears with the same factor in $4\Delta(\hat R_1)^2$ and $\Delta(\hat R_1-\hat R_2)^2$, meaning that one could perform a measurement which although not optimal would be pretty efficient. The same applies to the Schrödinger cat-like state $\ket{\psi_C}$ where one dominant term is $\Delta^2\omega_p^2$. 

\subsection{Phase space interpretation}

We now provide a geometrical interpretation of the previous results. If we consider that $\sigma_-\gg \sigma_+$ in the case of a Gaussian state or $\Delta\gg \sigma_+$ in the case of a Schr\"odinger cat-like state, the projection of the Wigner function on the plane corresponding to collective minus variables  $(\tau_-, \phi_-)$  is the one presenting a relevant phase space structure. Thus, it would be interesting to consider, as in the case of translations, that these states are manipulated using operators acting on modes associated to this collective variable. A naïve guess would then trying to apply the rotation operator $\hat R_-$. However it comes with many difficulties. Indeed it first poses an experimental problem, since this rotation corresponds to a non-local action which would be very hard to implement. In addition, the HOM is not able to measure such evolution. Finally, it turns out that this is not the operator with the greatest QFI. This fact can be understood by taking a more careful  look at the Wigner function of the considered states. The Wigner function for separable states can be factorized as the product of two Wigner functions defined in variables plus and minus, and we have that $W_+$ is the Wigner function of a Gaussian state centered at $\omega_p$ (corresponding to the situation (a) in Figure~(\ref{fig: rotation states}). As for $W_-$, it is either the Wigner function of a Gaussian state or the one associated to a superposition of two Gaussian states centered around zero (corresponding to the situation (b) and (c) in Figure~\ref{fig: rotation states}). The QFI increases with the distance of the states to the rotation point. For this reason, states $\ket{\psi_G}$ and $\ket{\psi_C}$ under a rotation using $\hat R_-$, do not lead to a high QFI.\\

A higher QFI is obtained using rotations around a point which is far away from the center of the state. In this case, the QFI displays a term which is proportional to the distance from the center of rotation squared divided by the width of the state squared. Both terms $\omega_p^2\sigma_-^2$ and $\Delta^2\omega_p^2$ which were dominant in the expression of the variance of $\hat R_1$ and $\hat R_1-\hat R_2$ can be interpreted as such. This means that the rotation $\hat R_1$, whose action is not easily seen in the variables plus and minus, can be interpreted as a rotation which moves $W_-$ around the distance $\omega_p$ from the origin of the TF phase space ($\omega=0$).

For both states then, the main numerical contribution to the QFI comes from a classical effect, related to the intrinsic resolution associated to the central (high) frequency of the field. In general, in phase space rotations, both in the quadrature and in the TF configuration, the distance from the phase space origin plays an important role. While in the quadrature configuration this distance has a physical meaning that can be associated both to the phase space structure and to the number of probes. In the case of TF phase space, the distance from the origin and the phase space scaling are independent. In particular, the distance from the origin can be considered as a classical resource that plays no role on the scaling with the number of probes. 

\subsection{A discussion on scaling properties of rotations}

The different types of FT phase space rotations have different types of interpretation in terms of scaling. The combined rotations of the type $\hat R_1 \pm \hat R_2$, for instance, can be generalized to an $n$ photon set-up through operators as $\hat {\cal R}=\sum_i^n \alpha_i \hat R_i$, with $\alpha_i=\pm 1$. In this situation, we have that rotation operators are applied individually and independently to each one of the the $n$ photons. In this case, we can expect, in first place, a collective (classical) effect, coming simply from the fact that we have $n$ probes (each photon). In addition, it is possible to show that a Heisenberg-like scaling can be obtained by considering states which are maximally mode entangled in a mode basis corresponding to the eigenfunctions of operators $\hat R_i$. Indeed, for each photon (the $i$-th one), we can define a mode basis such that $\hat R_i \ket{\phi_{k}}_i=(k+1/2)\ket{\phi_{k}}_i$, with $\ket{\phi_{k}}_i=\frac{1}{\sqrt{2^{k} {k}!}} \frac{1}{\pi^{1/4}}\int d\omega e^{-\frac{ \omega^2}{2}}H_{k}(\omega)\ket{\omega}_i$ with $H_{k}(\omega)$ being the $k$-th Hermite polynomial associated to the $i$-th photon. For a maximally entangled state in this mode basis, \ie, a state of the type $\ket{\phi}=\sum_{k=0}^{\infty} A_{k}\bigotimes_{i=1}^n \ket{\phi_{k}}_i$, (where we recall that the subscript $i$ refers to each photon and $k$ to the rotation eigenvalues) the $\hat {\cal R}$ eigenvalues behave as random classical variables and we can show that the QFI scales as $n^2$. 

As for rotations of the type $\hat R_{\pm}$, they cannot be decomposed as independently acting on each photon, but consist of entangling operators that can be treated exactly as $\hat R_1$ and $\hat R_2$ but using variables $\omega_{\pm}=\omega_1 \pm \omega_2$ instead of $\omega_1$ and $\omega_2$. We can also compute the scaling of operators as $\hat J = \sum_{\Omega_{\beta}} \hat R_{\Omega_{\beta}}$ where $\Omega_{\beta}=\sum_i^n \alpha_i \omega_i$, $\alpha_i=\pm 1$ and $\beta$ is one of the $2^{n-1}$ ways to define a collective variable using the coefficients $\alpha_i$.  For such, we can use the same techniques as in the previous paragraph but for the collective variables $\Omega_{\beta}$. Nevertheless, the experimental complexity of producing this type of evolution and the entangled states reaching the Heisenberg limit is such that we'll omit this discussion here.

\section{Conclusion}

We have extensively analyzed a quantum optical set-up, the HOM interferometer, in terms of its quantum metrological properties.  We provided a general formula for the coincidence probability of this experiment which led to a general formula for the associated FI.  We used this formula to analyze different types of evolution and showed when it is possible to reach the QFI in this set-up. In particular, we made a clear difference between collective quantum effects that contribute to a better than classical precision scaling and classical only effects, associated to single mode spectral properties. We then briefly discussed the general scaling properties of the QFI associated to the studied operators. 

Our results provide a complete recipe to optimize the HOM experiment with metrological purposes. They rely on the symmetry properties of quantum states that are revealed by the HOM interferometer. An interesting perspective is to generalize this type of reasoning for different set-ups where different symmetries play a role on the measurement outputs. 

\section*{Acknowledgements}

The French gouvernement through the action France 2030 from Agence Nationale de la Recherche, reference ``ANR-22-PETQ-0006" provided financial support to this work. We thanks Nicolas Fabre for fruitful discussions and comments on the manuscript. 
\bibliographystyle{unsrturl} 
\bibliography{refs}

\appendix
\addtocontents{toc}{\protect\setcounter{tocdepth}{1}} 
\onecolumngrid
\section{Time frequency formalism}\label{annex: tf formalism}
In quantum mechanics, light is described with the help of modes \cite{fabre_modes_2020}, representing the various physical properties a photon can have: frequency, position, spectral shape, wave vector, polarization... Mathematically we associate to each mode $\alpha$ a creation and annihilation operators $\hat{a}^\dagger_\alpha$ and $\hat{a}_\alpha$ which satisfy the familiar bosonic commutation relation $[\hat{a}_\alpha,\hat{a}^\dagger_\beta]=\delta_{\alpha,\beta}$. The quantum states are then obtained by acting with the creation operators on the vacuum $\ket{\text{vac}}$, which can be interpreted as adding a photon in the corresponding mode.\\

In time frequency continuous variables we look at modes parameterized by the frequency \cite{fabre_time_2022}. We will thus adapt the terminology: for us a mode will correspond to all physical parameter needed to describe a photon excluding the frequency (position, wave vector, polarization...). In the following we will look at interferometers, and thus the parameter $\alpha$ will describe in which arm the photon is propagating. We will thus describe single photon states in a given mode $\alpha$ with frequency $\omega$ with the help of a creation operator acting on the vacuum state: $\hat{a}_\alpha^\dagger(\omega)$. In this situation the commutation relation is written as 
 \begin{equation}
     [\hat{a}_\alpha(\omega),\hat{a}_\beta^\dagger(\omega')]=\delta(\omega-\omega')\delta_{\alpha,\beta},
 \end{equation}
 the other commutation relations (between two creation or two annihilation operators) vanishing. It's useful to introduce the conjugated temporal variable $t$, by the use of the Fourier transform:
 \begin{equation}
     \hat{a}_\alpha(t)=\frac{1}{\sqrt{2\pi}}\int d\omega \hat{a}_\alpha(\omega)e^{-i\omega t}.
 \end{equation}
 
 We can verify that the creation and annihilation operators in the temporal domain verify the same commutation relation as the one in the spectral domain: $[\hat{a}_\alpha(t),\hat{a}_\beta^\dagger(t')]=\delta(t-t')\delta_{\alpha,\beta}$.

\subsubsection{States in time-frequency variables}
The creation operators allow to define general single photon states on a single mode via:
\begin{equation}
    \ket{\psi}=\int d\omega S(\omega)\hat{a}^\dagger(\omega)\ket{\text{vac}}=\int d\omega S(\omega)\ket{\omega}.
\end{equation}

The spectrum $S(\omega)$ is the Fourier transform of the time of arrival distribution and it can be recovered from the state $S(\omega)=\braket{\omega}{\psi}$. If we are interested in a collection of $n$ single photons states in $n$ different modes, we can work with the state:
\begin{equation}
    \ket{\psi}=\int  d\omega_1\cdots  d\omega_n F(\omega_1,\cdots,\omega_n)\hat{a}_1^\dagger(\omega_1) \cdots \hat{a}_n^\dagger (\omega_n)\ket{\text{vac}}=\int  d\omega_1\cdots  d\omega_n F(\omega_1,\cdots,\omega_n)\ket{\omega_1,\cdots,\omega_n},
\end{equation}
where the spectral function $F$ is normalised to one: $\int \abs{F(\omega_1,\omega_2}^2d\omega_1 d\omega_2=1$.

\subsubsection{Time-frequency operators}
We can introduce two very useful operators as follows:
\begin{align}
    \hat{t}_\alpha=\int\limits dt ~t \hat{a}_\alpha^\dagger(t)\hat{a}_\alpha(t) && 
    \hat{\omega}_\alpha=\int d\omega ~\omega \hat{a}_\alpha^\dagger(\omega)\hat{a}_\alpha(\omega).
\end{align}

The fundamental property of these operators is the fact that they verify the familiar commutation relation on the subspace of single photons:
\begin{equation}
    \label{eq: commutator w t}
    [\hat{\omega}_\alpha,\hat{t}_\alpha]=i.
\end{equation}
More precisely, we have the general result:
\begin{equation}
    \label{eq: commutator w t n photon}
    [\hat{\omega}_\alpha,\hat{t}_\alpha]=i\int\limits_{-\infty}^\infty d\omega \hat{a}_\alpha^\dagger(\omega)\hat{a}_\alpha(\omega)=i \hat{N}_\alpha,
\end{equation}
where the operator $\hat{N}_\alpha$ count the number of photon operator in the mode $\alpha$.\\

The action of the both operators $\hat \omega$ and $\hat t$ can be computed on the JSA and we have:
\begin{align}
    \hat \omega: S(\omega)\mapsto \omega S(\omega) && \hat t: S(\omega)\mapsto -i\partial_\omega S(\omega).
\end{align}

\section{Appendix: Derivation of equations (\ref{eq: pc}) and (\ref{eq: fi})}\label{annex: proofs}
\subsubsection{Equation (\ref{eq: pc})}
To show equation (\ref{eq: pc}) we start with the state before the BS:
\begin{equation}
    \hat{U}\ket{\psi}=\int d\omega_1 d\omega_2 F(\omega_1,\omega_2) \ket{\omega_1,\omega_2}.
\end{equation}
The usual balanced BS relation reads:
\begin{equation}
    \ket{\omega_1}_1\ket{\omega_2}_2\mapsto\frac{1}{2}\Big[\ket{\omega_1}_1\ket{\omega_2}_1-\ket{\omega_1}_1\ket{\omega_2}_2+\ket{\omega_1}_2\ket{\omega_2}_1-\ket{\omega_1}_2\ket{\omega_2}_2\Big].
\end{equation}
To be able to use it, we introduce two mode changing operators $\hat T_1$ and $\hat T_2$ defined by:
\begin{align}
    \hat T_1\ket{\omega_1}_1\ket{\omega_2}_2=\ket{\omega_1}_1\ket{\omega_2}_1 && \hat T_2\ket{\omega_1}_1\ket{\omega_2}_2=\ket{\omega_1}_2\ket{\omega_2}_2.
\end{align}
With these definition the BS splitter relation is equivalent to applying the operator:
\begin{equation}
    \frac{1}{2}(\hat T_1-\hat{ \mathds{1}}+\hat S-\hat T_2),
\end{equation}
where $\hat S$ is the swap operator, defined as $\hat S \ket{\omega_1, \omega_2}=\ket{\omega_2,\omega_1}$ So the state coming out of the BS is:
\begin{equation}
    \ket{\psi_{\text{out}}}=\frac{1}{2}\int d\omega_1 d\omega_2 F(\omega_1,\omega_2)\Big[\hat T_1 \hat U-\hat U+\hat S\hat U-\hat T_2\hat U\Big]\ket{\omega_1,\omega_2}.
\end{equation}
If we do selection on coincidence, we only keep the part of the state with one photon in each mode. We get the state:
\begin{subequations}
\begin{align}
    \ket{\psi_{\text{fin}}}&=\frac{-1}{2}\int d\omega_1 d\omega_2 F(\omega_1,\omega_2)\Big[\hat U -\hat S\hat U\Big]\ket{\omega_1,\omega_2}\\
    &=\frac{1}{2}\Big[\hat S\hat U-\hat U\Big]\ket{\psi}.
\end{align}
\end{subequations}
We can finally compute the coincidence probability by taking the norm square of $\ket{\psi_{\text{fin}}}$:
\begin{subequations}
\begin{align}
    P_c&=\braket{\psi_{\text{fin}}}\\
    &=\frac{1}{4}\bra{\psi}\Big[\hat U^\dagger-\hat U^\dagger\hat S\Big]\Big[\hat U-\hat S\hat U\Big]\ket{\psi}\\
    &=\frac{1}{4}\bra{\psi}\Big[\underbrace{\hat U^\dagger\hat U}_{=\mathds{1}}-2\hat U^\dagger\hat S\hat U+\underbrace{\hat U^\dagger \hat S\hat S\hat U}_{=\hat U^\dagger\hat U=\mathds{1}} \Big]\ket{\psi}\\
    &=\frac{1}{2}\Big[1- \bra{\psi}\hat U^\dagger \hat S\hat U \ket{\psi}\Big].
\end{align}
\end{subequations}

\subsubsection{Equation (\ref{eq: fi})}
The expression for $\mathcal{Q}$ is a direct consequence of the expression of the QFI for pure state. \\
The proof of the expression of $\mathcal{F}$ is a little bit more involved. We have to compute:
\begin{equation}
    FI(\kappa)=\frac{1}{P_c}\left(\frac{\partial P_c}{\partial \kappa}\right)^2+\frac{1}{P_a}\left(\frac{\partial P_a}{\partial \kappa}\right)^2.
\end{equation}

We have seen the expression of the (anti)-coincidence probability $P_c$ and $P_a$ that depends on $\bra{\psi}\hat U^\dagger \hat S \hat U\ket{\psi}$. If we make the assumption that the state $\ket{\psi}$ is either symmetric or anti-symmetric we known that we have: $\bra{\psi}\hat U^\dagger \hat S \hat U\ket{\psi}=\pm\bra{\psi}\hat U^\dagger \hat S \hat U\hat S \ket{\psi}=\bra{\psi}\hat V(\kappa)\ket{\psi}$ where we denote  $\hat{V}(\kappa)=\hat{U}^\dagger \hat S\hat{U} \hat S=e^{i\kappa \hat{H}}e^{-i\kappa\hat S\hat H\hat S}$. We first start by expanding this scalar product up to the second order in $\kappa$, using the short hand notation $\expval{\cdot}=\bra{\psi}\cdot\ket{\psi}$.

\begin{subequations}
\begin{align}
    \bra{\psi}\hat V(\kappa)\ket{\psi}&=\expval{e^{i\kappa\hat H}e^{-i\kappa\hat S\hat H\hat S}}\\
    &\simeq \expval{\Big(1+i\kappa\hat H-\frac{\kappa^2}{2}\hat H^2\Big)\Big(1-i\kappa\hat S\hat H\hat S-\frac{\kappa^2}{2}(\hat S\hat H\hat S)^2\Big)}\\
    &=\expval{1+i\kappa\hat H-i\kappa\hat S\hat H\hat S-\frac{\kappa^2}{2}\hat H^2-\frac{\kappa^2}{2}(\hat S\hat H\hat S)^2+\kappa\hat H\hat S\hat H\hat S}
    \intertext{Since the state $\ket{\psi}$ is (anti)-symmetric, for any operators $\hat G$, we have $\expval{\hat S\hat G}=\pm\expval{\hat G}=\expval{\hat G\hat S}$, which allows some simplifications.}
    &=1-\frac{\kappa^2}{2}\Big(\expval{\hat H^2}+\expval{(\hat S\hat H\hat S)^2}-\expval{\hat H\hat S\hat H\hat S}-\expval{\hat S\hat H\hat S\hat H}\big)\\
    &=1-\frac{\kappa^2}{2}\expval{(\hat H-\hat S\hat H\hat S)^2}\\
    &=1-\frac{\kappa^2}{2}\Delta(\hat H-\hat S\hat H\hat S)^2
    \intertext{Since thanks to the symmetry of $\ket{\psi}$, $\expval{\hat H-\hat S\hat H\hat S}=\expval{\hat H-\hat H\hat S^2}=0$}\notag
\end{align}
\end{subequations}

By defining $\hat G=\hat H-\hat S\hat H\hat S$ it remains to compute the FI:

\begin{subequations}
\begin{align}
    FI(\kappa=0)&=\frac{1}{P_c}\left(\frac{\partial P_c}{\partial \kappa}\right)^2+\frac{1}{P_a}\left(\frac{\partial P_a}{\partial \kappa}\right)^2\\
    &=\frac{1}{4P_c}\left(\kappa\Delta(\hat{G})^2\right)^2+\frac{1}{4P_a}\left(\kappa\Delta(\hat{G})^2\right)^2\\
    &=\frac{\kappa^2\Delta(\hat{G})^4}{4}\left(\frac{1}{P_c}+\frac{1}{P_a}\right)\\
    &=\frac{\kappa^2\Delta(\hat{G})^4}{4}\frac{P_a+P_c}{P_cP_a}\\
    &=\frac{\kappa^2\Delta(\hat{G})^4}{4}\frac{4}{\left(1+\bra{\psi}\hat{V}(\kappa)\ket{\psi}\right)\left(1-\bra{\psi}\hat{V}(\kappa)\ket{\psi}\right)}\\
    &=\kappa^2\Delta(\hat{G})^4\frac{1}{1-\bra{\psi}\hat{V}(\kappa)\ket{\psi}^2}\\
    &=\kappa^2\Delta(\hat{G})^4\frac{1}{\kappa^2\Delta(\hat{G})^2}\\
    &=\Delta(\hat{G})^2
\end{align}
\end{subequations}

It is interesting to note that the computation of the Fisher information is singular. Indeed for the HOM interferometer around $\kappa=0$ the derivative of the probabilities vanishes $\partial_\kappa P_{c,a}=0$, while one of the two probability ($P_c$ if the state is symmetric or $P_a$ if its anti-symmetric) is also equal to zero. We thus obtain here the FI at zero by computing it at $\kappa\neq 0$ and taking the limit. As a result we see that the FI is proportional to the second derivative of the coincidence probability. This means  that for such a measurement what is important is the curvature of the probability peak/dip.

\section{Appendix: Details on the computation of the various variances}\label{annex variance computations}
To compute explicitly the various variances of this paper on the two states $\ket{\psi_G}$ and $\ket{\psi_G}$ one can note that since these states are separable in the variables $\omega_\pm$, if we consider two operators $\hat H_+$ and $\hat H_-$ which are respectively functions of $\hat\omega_+$ and $\hat t_+$ or $\hat\omega_-$ and $\hat t_-$ we have: $\expval{\hat H_+\hat H_-}=\expval{\hat H_+}\expval{\hat H_-}$. Where for a fixed state $\ket{\psi}$, $\expval{\hat H}= \bra{\psi}\hat H\ket{\psi}$.\\

In order to compute any variance, one only has to compute some expectation values. By expanding and using the independence property from above, one only need to compute as building block expectation value of the form: $\expval{\hat\omega_\pm^k\hat t_\pm ^l}$. Indeed we can use the commutation relation to reorder  any product such that the frequency operators are on the left of the time operators. One has to pay attention that due to the choice of normalisation in the definition of $\hat\omega_\pm=\hat\omega_1\pm\hat\omega_2$ and $\hat t_\pm=\hat t_1\pm \hat t_2$ we have $[\hat \omega_\pm,\hat t_\pm]=2i$. Such expectation values can be obtained systematically using a software (here we used Mathematica), we have the following values:

\begingroup
\setlength{\tabcolsep}{10pt} 
\renewcommand{\arraystretch}{2} 
\begin{table}[H]
    \centering
    \begin{tabular}{|c|c|c|c|}
        \hline
       Operator & Variable $+$ & Variable $-$ for $\ket{\psi_G}$& Variable $-$ for $\ket{\psi_C}$ \\
        \hline
        $\hat{\omega}$ & $\omega_p$ & $0$ & $0$\\
        $\hat{\omega}^2$ & $\omega_p^2+\sigma_+^2$ & $\sigma_-^2$ & $\sigma_-^2+\frac{1}{4}\Delta^2$\\
        $\hat{\omega}^3$ & $3\sigma_+^2\omega_p+\omega_p^3$ & $0$ & $0$\\
        $\hat{\omega}^4$ & $3\sigma_+^4+6\sigma_+^2\omega_p^2+\omega_p^4$ & $3\sigma_-^4$ & $3\sigma_-^4+\frac{3}{2}\sigma_-^2\Delta^2+\frac{1}{16}\Delta^4$\\
        \hline
        $\hat{t}$ & $0$ & $0$& $0$\\
        $\hat{t}^2$ & $\frac{1}{\sigma_+^2}$ & $\frac{1}{\sigma_-^2}$& $\frac{1}{\sigma_-^2}$\\
        $\hat{t}^3$ & $0$ & $0$& $0$\\
        $\hat{t}^4$ & $\frac{3}{\sigma_+^4}$ & $\frac{3}{\sigma_-^4}$& $\frac{3}{\sigma_-^4}$\\
        \hline
        $\hat{\omega}\hat{t}$ & $i$ & $i$& $i$\\
        $\hat{\omega}^2\hat{t}$ & $2i\omega_p$ & $0$ & $0$\\
        $\hat{\omega}\hat{t}^2$ & $\frac{\omega_p}{\sigma_+^2}$ & $0$& $0$\\
        $\hat{\omega}^2\hat{t}^2$ & $\frac{\omega_p^2}{\sigma_+^2}-1$ & $-1$ & $\frac{\Delta^2}{4\sigma_-^2}-1$ \\
        \hline
    \end{tabular}
    \caption{Expectation values of the various product of plus and minus operators on the states $\ket{\psi_G}$ and $\ket{\psi_C}$.}
    \label{tab:expectation values plus minus}
\end{table}
\endgroup

 \end{document}